# Comment on "Damping Force in the Transit-time Method of Optical Stochastic Cooling"

In Ref. [1] the author claimed, that he clarifies a specific aspect of a transient time method of optical stochastic cooling [2]. The center point of his concept associated with basic transformation of energy in a kicker undulator described by the formula

$$\delta_i^{(2)} = \delta_i^{(1)} - G \cdot \sin(kR_{56}\delta_i^{(1)}) \quad (1)$$

where $\delta_i^{(1)}$, $\delta_i^{(2)}$ is a particle's relative energy before and after the cooling insertion respectively, $k = 2\pi/\lambda$, $\lambda$ is the optical wavelength, $R_{56}$ is the time-of-flight characteristics of the cooling insertion, and the $G$ is a damping coefficient. Author uses equation (1) even when the particle manifests substantial slippage with respect to its own wavelet, emitted in a pickup undulator and optically amplified. One can see that if the phase delay $kR_{56}\delta_i^{(1)}$ has an addition factor $2\pi n$ ($n=1,2,3,\ldots$) the term $G \cdot \sin(kR_{56}\delta_i^{(1)})$ does not change its value. Physically such delay means that the particle arrives at the kicker undulator with delay which corresponds to the integer number of the optical wavelengths. So according to formula (1) the kick will be the same.

In reality however, as the particle slides along the amplified signal of limited duration, the value of kick changes as well, Fig.1

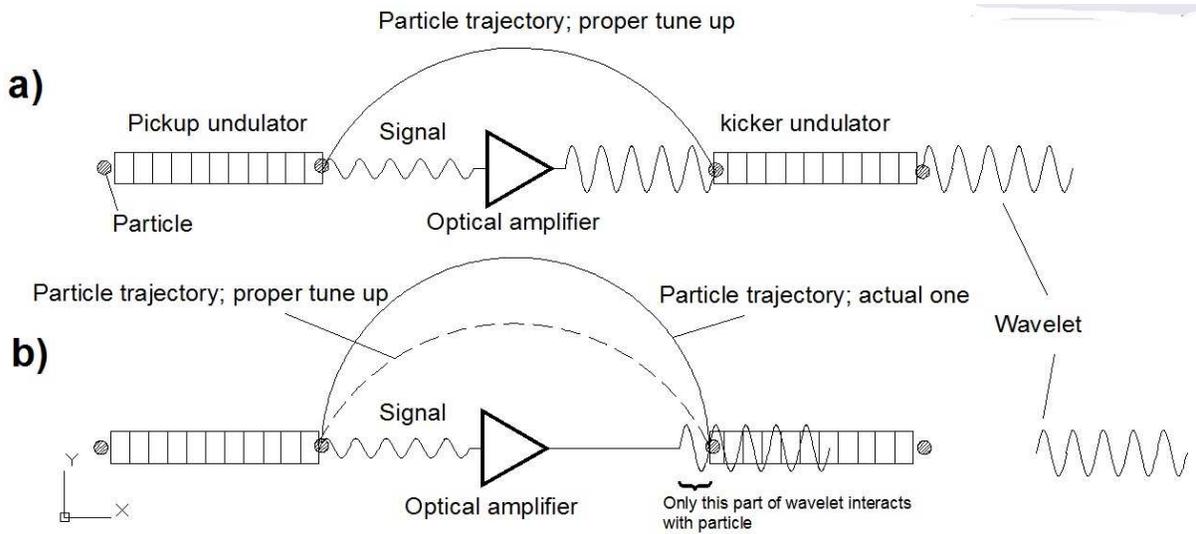

Figure 1. a) Proper tuned positions of particle and amplified signal wavelet; b) Actual positions of particle and signal wavelet. Systematic optical delay unit for adjustment of proper particle arrival time at the kicker undulator is not shown here. This unit compensates delay of optical signal in the amplifier also.

This can be seen from definition of $G$ [1]; simply less number of periods of the amplified signal act to the particle in a kicker undulator. With the other words, the factor $G$ is a function of delay also, so the correct formula looks like

$$\delta_i^{(2)} = \delta_i^{(1)} - G \cdot (1 - \frac{kR_{56}\delta_i^{(1)}}{N\lambda}) \cdot Sin(kR_{56}\delta_i^{(1)}),$$

where $N$ stands for the number of periods in a wavelet (number of periods of pickup undulator). For example, if the particle arrives at the kicker undulator and meets the only one period of the amplified wavelet, the cooling kick will be only $\sim 1/N$ of $G$ and so on, but formula (1) does not describe this at all. Namely this makes optimal phasing not as simple as it is described in the reference materials ([1], formula (3)). As the amplitude of the kick drops while the mismatch increased, fragmentation of phase space at larger amplitudes is reduced also.

As far as the fragmentation of the phase space, it was described (and explained) first in [3], so the claim of priority in illumination of this phenomenon is problematic also.

E.G.Bessonov
FIAN, Moscow, Russia

A.A.Mikhailichenko
Cornell U., CLASSE, Ithaca, U.S.A.

___________________________________